\documentclass{aa}  

\usepackage{graphicx}
\usepackage{txfonts}
\usepackage{hhline,tabu}
\usepackage{multirow}
\usepackage{tabularx}
\usepackage[]{hyperref}
\usepackage{upgreek}
\usepackage[dvipsnames]{xcolor}
\usepackage[normalem]{ulem}

\hypersetup{
    colorlinks = true,
    linkcolor = blue,
    anchorcolor = black,
    citecolor = blue,
    filecolor = cyan,
    menucolor = red,
    runcolor = cyan,
    urlcolor = purple
}

\usepackage{diagbox}
\usepackage{siunitx}
\newcommand{\orcid}[1]{\href{https://orcid.org/#1}{\protect\includegraphics[width=8pt]{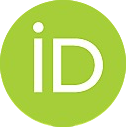}}}

\begin{document} 

\title{Cluster membership analysis with supervised learning and $N$-body simulations}   
\title{Cluster membership analysis with supervised learning and $N$-body simulations}   
   \author{A. Bissekenov\inst{1,2}\orcid{0009-0003-4608-2611},
          M. Kalambay\inst{3,4,2,5,6}\orcid{0000-0002-0570-7270},
          E. Abdikamalov\inst{7,2}\orcid{0000-0001-5481-7727},
          X. Pang \inst{1,8}\orcid{0000-0003-3389-2263},
          P. Berczik \inst{9,10,11,5}\orcid{0000-0003-4176-152X}
          \and
          B.~Shukirgaliyev\inst{3,4,7}\orcid{0000-0002-4601-7065}\fnmsep\thanks{Corresponding author: b.shukirgaliyev@hw.ac.uk}
          }

   \institute{
            Department of Physics, Xi'an Jiaotong-Liverpool University, 111 Ren'ai Road, Dushu Lake Science and Education Innovation District, Suzhou 215123, Jiangsu Province, P.R. China.
            \and
            Energetic Cosmos Laboratory, Nazarbayev University,
              53 Kabanbay Batyr Ave., 010000 Astana, Kazakhstan
         \and
            Heriot-Watt University Aktobe Campus, 263 Zhubanov Brothers Str, 030000 Aktobe, Kazakhstan 
         \and
            Heriot-Watt International Faculty, K.~Zhubanov Aktobe Regional University, 263 Zhubanov Brothers Str, 030000 Aktobe, Kazakhstan%\\
        \and
            Fesenkov Astrophysical Institute, 23 Observatory Str., 050020 Almaty, Kazakhstan         
        \and
            Faculty of Physics and Technology, Al-Farabi Kazakh National University, 71 Al-Farabi Ave, 050020 Almaty, Kazakhstan%\\
        \and
            Department of Physics, School of Sciences and Humanities, Nazarbayev University, 53 Kabanbay Batyr Ave., 010000 Astana
            \and
            Shanghai Key Laboratory for Astrophysics, Shanghai Normal University, 100 Guilin Road, Shanghai 200234, PR China
            \and
            Nicolaus Copernicus Astronomical Centre Polish Academy of Sciences, ul. Bartycka 18, 00-716 Warsaw, Poland
\and
Konkoly Observatory, HUN-REN Research Centre for Astronomy and Earth Sciences, Konkoly Thege Mikl\'os \' ut 15-17, 1121 Budapest, Hungary
            \and
Main Astronomical Observatory, National Academy of Sciences of Ukraine, 27 Akademika Zabolotnoho St., 03143 Kyiv, Ukraine
            }

   \date{Received 29 Feb 2024; accepted 29 Jul 2024}
 
  \abstract
   {Membership analysis is an important tool for studying star clusters. There are various approaches to membership determination, including supervised and unsupervised machine learning (ML) methods.}
   {We perform membership analysis using the supervised machine learning approach.}
   {We train and test our ML models on two sets of star cluster data: snapshots from $N$-body simulations and 21 different clusters from the {\it Gaia} Data Release 3 data.}
   {We explore five different ML models: Random Forest (RF), Decision Trees, Support Vector Machines, Feed-Forward Neural Networks, and K-Nearest Neighbors. We find that all models produce similar results, with RF showing slightly better accuracy. We find that a balance of classes in datasets is optional for successful learning. The classification accuracy depends strongly on the astrometric parameters. The addition of photometric parameters does not improve performance. We do not find a strong correlation between the classification accuracy and clusters' age, mass, and half-mass radius. At the same time, models trained on clusters with a larger number of members generally produce better results.}
   {}

   \keywords{(Galaxy:) open clusters and associations: general -- Methods: numerical
 -- Methods: data analysis }
    \titlerunning{Membership analysis with machine learning and $N$-body simulations}
   \authorrunning{Bissekenov et al.}
   \maketitle

%
%-------------------------------------------------------------------

\section{Introduction}

The research in open star clusters (OC) is essential in many areas of astronomy, including the formation and evolution of stars, Galactic dynamics, and star-formation history \citep{LL2003, onSFhistory2004, krumholz2019star}. The reason is that the stars in a cluster tend to have the same age and kinematics \citep{PZ+2010, Renaud2018, krumholz2019star}. To better estimate the fundamental parameters such as age, mass, size, and metallicity of star clusters, it is crucial to separate stars that are members from those that are in the field of the Galaxy \citep{Kharchenko2005, Kharchenko2012}. This process is usually referred to as membership analysis.  

Identifying member stars within open clusters presents a significant challenge due to their location within the Galactic disk \citep{Ascenso+2009, Kharchenko2012}. The subtle over-density generated by these clusters often becomes obscured by field stars \citep{Bland-HawthornGerhard2016}. Historically, membership analysis relied heavily on manual methodologies \citep[e.g.][]{Stock1956, Ruprecht+1981, PhelpsJanes1994, Chen+2003, Kharchenko2005, Kharchenko2012, Dias+2014, RAVE2017, Roser+2019, MeingastAlves2019, Lodieu+2019}. However, recent advancements in machine learning (ML)and big datasets provided by {\it Gaia} Data Releases \citep{gaiadr1,gaiamission,gaiadr2,gaiaedr3,gaiadr3} have revolutionized this process \citep{Olivares+2023}. Numerous studies now leverage unsupervised learning techniques to automate and enhance the determination of cluster membership \citep[][and more]{gao2014membership, cantat2018gaia, gao2018memberships, tang+2019, LiuPang2019, Castro-Ginard+2020, agarwal2021ml, Noormohammadi+2023, HuntReffert2023}. Generally, one can use supervised and unsupervised ML methods \citep{survey_sup_unsup}. While the supervised approach is based on training on labeled data, the unsupervised method does not rely on labeled data \citep{survey_sup_unsup}. 

There are several types of unsupervised learning methods \citep{bishop2006pattern}, including clustering, self-organized map algorithms, and probabilistic models \citep{baccao2005self}. Among clustering methods, the density-based algorithms gain wide use \citep{gao2014membership, tucio2023investigation, ghosh2022membership}. There are two versions of this method: density-based spatial clustering of applications with noise \citep[DBSCAN,] []{ester1996density}, and hierarchical density-based spatial clustering of applications with noise \citep[HDBSCAN,][]{hdbscan}. DBSCAN works by assigning (i.e., using hyperparameters) the radius and the minimal number of stars \citep{census2021}. If the numbers of actual stars exceed these values, this region is classified as a cluster \citep{gao2014membership}. HDBSCAN works similarly but extends DBSCAN by constructing a cluster hierarchy based on minimum spanning trees \citep{census2021}. This allows handling clusters of varying shapes and densities \citep{hdbscan}. \citet{gao2014membership} used DBSCAN on the 3D kinematic features of the NGC 188 cluster and identified 1,504 member star candidates. \citet{tucio2023investigation} and \citet{ghosh2022membership} used HDBSCAN to separate the member stars of NGC 2682 and NGC 7789 from field stars. The UPMASK code combines principle component analysis and clustering algorithms such as k-means clustering implemented in R language \citep{upmask2015}. pyUPMASK is a Python implementation of the same method \citep{pera2021pyupmask}. StarGO \citep{stargo2018} is an unsupervised self-organized map algorithm, initially built to find halo stars in the Galaxy. \citet{tang+2019} used it for dimensionality reduction and membership determination by using 5D kinematic data of stars. Later, this technique was used to find new OCs in the solar neighborhood \citep{LiuPang2019, pang2020different, pang65membership}. Additionally, a probabilistic model such as the Gaussian Mixture Model (GMM) was used in the work of \citet{edgmm2021}, where they applied it to 426 OCs. 

There were also several attempts at supervised learning on membership analysis. A combination of unsupervised learning with supervised model methods was employed, such as random forest (RF) with GMM \citep{gao2018machine, gao2019investigation, gao2019praesepe, mahmudunnobe2021membership,  jadhav2021, das2023membership, guido2023}, $K$-nearest neighbors (KNN) with GMM \citep{agarwal2021ml, deb2022ensemble}. As for the supervised learning-only approach, \citet{groeningen2023} used Deep Sets Neural Networks on 167 OCs from the {\it Gaia} DR2 and eDR3 data based on the catalog of \citet{can_gau2020}.

However, membership analysis performed using different methods in previous studies shows limited agreement with each other. For example, \citet{Bouma+2021} reports that for open cluster NGC~2516, 25\% of \citet{kounkel2019untangling} labels, 41\% of \citet{meingast2021extended} labels, and 68\% of \citet{cantat2018gaia} labels overlap with each other. 

In this work, we perform membership analysis using supervised learning on data from simulations and observation for both training and testing. For the former, we use $N$-body simulations from \citet{bek2021dehnen}. The advantage of using simulation data is that we can determine the memberships of stars based on physical conditions in the simulation. We use the {\it Gaia} DR3 data for the observational data and label the cluster members based on findings of \citet{LiuPang2019} and \citet{pang65membership}. 

This paper is organized as follows. Section \ref{sec:method} described methodology. Section \ref{sec:results} presents results, and Section \ref{sec:conclusion} provides conclusions. 
   
\section{Method}
\label{sec:method}

\subsection{Data}
\label{sec:data}

We train our ML models using data from $N$-body simulations and {\it Gaia} DR3 observations. {When we use the simulation data, we consider} eight observable parameters to describe stars. These are five astrometric parameters: right ascension $\alpha$, declination $\delta$, two proper motions along these directions $\mu_{\alpha}$ and $\mu_{\delta}$, and parallax $\pi$; two photometric parameters: apparent magnitude $m_G$ and color index $G_{BP}-G_{RP}$ in {\it Gaia} bands \citep{apellaniz2018reanalysis}; and the radial velocity $\upsilon_r$. To study the impact of these parameters, we explore different combinations of these parameters in our training and testing. For our default combination, which we call combination 1, we use a 5-parameter family ($\alpha$, $\delta$, $\mu_\alpha$, $\mu_\delta$, $\pi$). For parameter combination 2, we use ($\alpha$, $\delta$, $m$, $G_{BP}-G_{RP}$. For combinations 3, 4, and 5, we use ($\alpha$, $\delta$, $\mu_\alpha$, $\mu_\delta$, $m$, $G_{BP}-G_{RP}$), ($\alpha$, $\delta$, $\mu_\alpha$, $\mu_\delta$, $\pi$, $\upsilon_r$), and ($\alpha$, $\delta$, $\mu_\alpha$, $\mu_\delta$, $\pi$, $m$, $G_{BP}-G_{RP}$). {Not all of these data combinations are available in the {\it Gaia} catalog. For this reason, when we test and train using the {\it Gaia} data, we use only the default combination (the combination 1).} See Table \ref{tab:feature} for a summary. 

\subsubsection{$N$-body simulations} \label{sec:nbodymethod}

Our study is partially based on applications of supervised learning on $N$-body simulation data that covers dynamic evolution from the gas expulsion to complete dissolution \citep{shukirgaliyev2017impact,bek2021dehnen}. The initial conditions are based on the \citet{parmentier2013local} model. We use $N$-body simulation models of \citet{bek2021dehnen} with Plummer density profile at the time of instantaneous gas expulsion. Those are clusters of $10^4$ stars formed with three different global star-formation efficiencies (SFEs) of $17\%,$ $20\%,$ and $25\%.$ \citet{bek2021dehnen} performed nine simulations per model with different randomization. We use only two randomizations of position and mass labeled as `11' and `22' of \citet{bek2021dehnen}. We did not use model clusters with the lowest SFE of $15\%$ because they dissolve soon due to their low star count in the aftermath of gas expulsion.

$N$-body simulations use Cartesian coordinates with origin at the Galactic center \citep{shukirgaliyev2019violent}. We then place the cluster at a distance of 150 pc from the Sun as shown in the \citet{kalambay2022mock-mukha}. We transfer the data to equatorial (International Celestial Reference System - ICRS) coordinates using the Astropy package \citep{robitaille2013astropy} to derive five astrometric parameters $\alpha$, $\delta$, $\mu_\alpha$, $\mu_\delta$, and $\pi$. Using effective temperature, mass, metallicity, and luminosity of each star available in $N$-body simulations snapshots, we calculate the absolute magnitude corresponding to the {\it Gaia} DR2 \citep[$G$, $G_{BP}$, $G_{RP}$, ][]{apellaniz2018reanalysis} bands using the method described in \citet{chen2019ybc}. Finally, with the knowledge of the parallaxes of all stars and their $G$, we calculate their apparent magnitudes ($m_G$) using
\begin{equation}
m_{G}=G-5(\log{\pi}+1).
\label{Mag-eq}
\end{equation} 
{We exclude the dim stars with $m_G$ < $21$~mag in the {\it Gaia} DR2 catalog from our analysis \citep{gaiadr2}}.

To make our simulation data resemble actual astronomical observations, we generate field stars of the galactic plane with the {\tt Galaxia} code for producing a synthetic galaxy model \citep{sharma2011galaxia}. We use the modified version by \citet{rybizki2019galaxia_wrap} to fit {\it Gaia} DR2. We then place our cluster inside these stars. We label stars within Jacobi radii \citep{Just+2009} as member stars, and those outside are labeled as non-members. These non-members were member stars in earlier $N$-body simulation snapshots. The field stars obtained from {\tt Galaxia} are also labeled non-members, though some may occur within the Jacobi radius of the cluster.

This work considers a circular field of view around the cluster center, covering about 600 square degrees. Within this region, we can have both types of stars of the simulated cluster that are gravitationally bound and unbound. We remove the tidal tail stars that are located outside this region. Nevertheless, some of the tail stars may appear in front of or behind the cluster by being projected onto the celestial sphere. Before applying ML, we move the origin of our equatorial coordinate system to the cluster center. We find the cluster center using the observational coordinates and proper motions using an iterative search of the density center. {In this algorithm, we find the average coordinates ($\alpha$ and $\delta$ in our case) of all stars in our set and shift the coordinate origin to this location. Then, we repeat this process, considering stars within $80\%$ of the maximum radius in the previous set until we reach the minimum threshold of 150 stars.}
 
\begin{table}
    \caption{Sets of feature combinations used in our classification analysis.}
    \begin{tabular}{|c|llllllll|}
    \hline
    \textbf{Set} & \multicolumn{8}{|c|}{\textbf{Features}}\\
    \hline
    1 &$\alpha$, & $\delta$,& $\mu_\alpha$,& $\mu_\delta$,& $\pi$& \ - & \ \ \ \ \ \ \ \ - &\ -\\
    \hline
    2 &$\alpha$,& $\delta$,&\ - &\ - & - &$m$,& $G_{BP}-G_{RP}$&\ -\\
    \hline
    3 & $\alpha$,& $\delta$,& $\mu_\alpha$,& $\mu_\delta$,& - & $m$,& $G_{BP}-G_{RP}$&\ -\\
    \hline
    4 & $\alpha$,& $\delta$,& $\mu_\alpha$,& $\mu_\delta$,& $\pi$,&\ -&\ \ \ \ \ \ \ \ - & $\upsilon_r$\\
    \hline
    5 & $\alpha$,& $\delta$,& $\mu_\alpha$,& $\mu_\delta$,& $\pi$,& $m$,& $G_{BP}-G_{RP}$&\ -\\
    \hline
    \end{tabular}
    \label{tab:feature}
\end{table}

\begin{figure}
	\includegraphics[width=0.8\columnwidth]{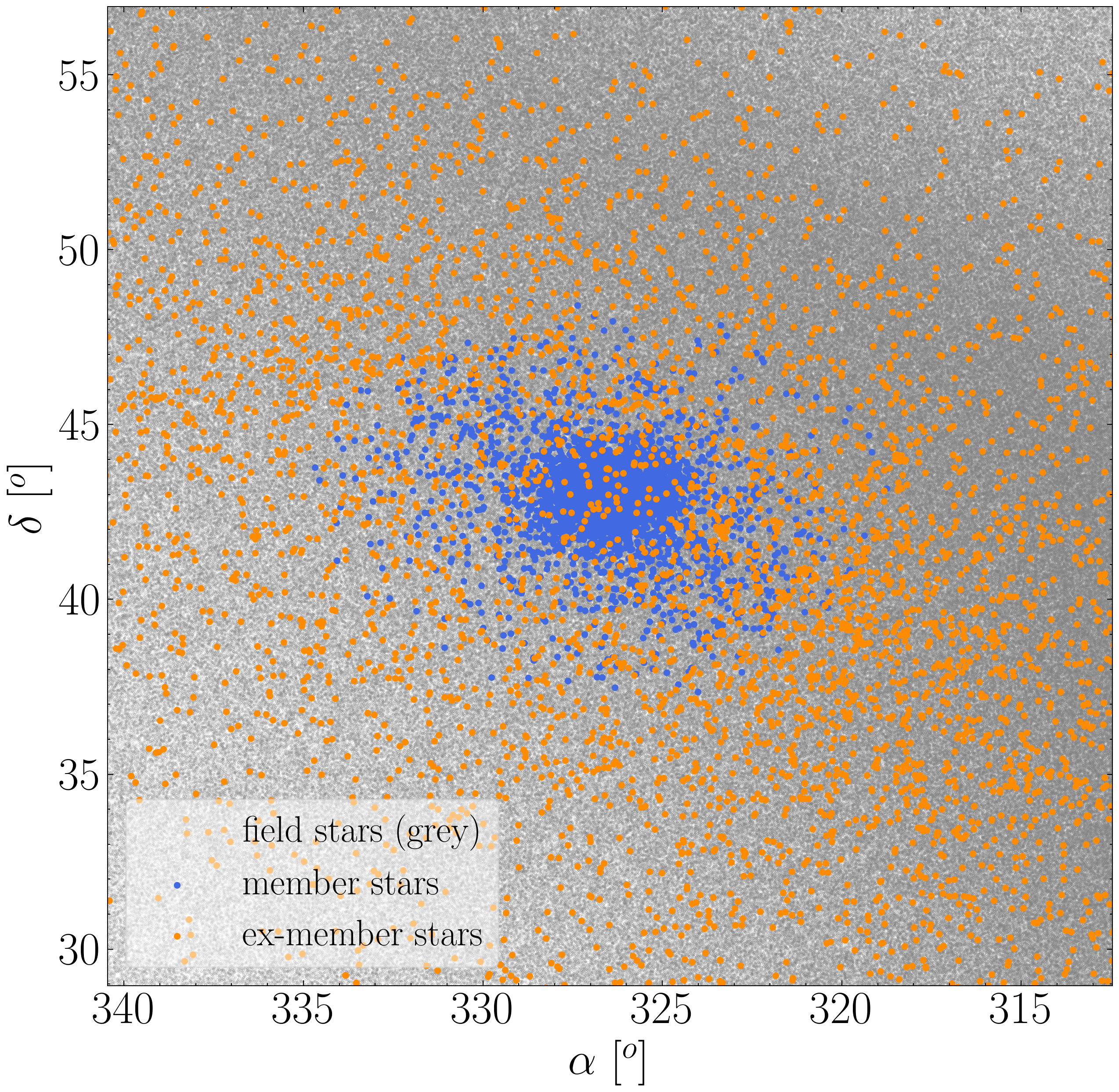}
    \caption{Stars from $N$-body simulation with SFE of $17\%$ at 20 Myr snapshot with generated field stars projected onto the sky. Blue, orange, and grey points show the member, non-member, and galactic field stars.}
    \label{fig:field_example}
\end{figure}

\subsubsection{{\it Gaia} data}
\label{gaia_data}

Besides using data from $N$-body simulations, we also perform training and testing on clusters from the {\it Gaia} DR3 data \citep{gaiadr3}. From now on, we refer to {\it Gaia} DR3 as {\it Gaia} data. We train three supervised ML models on Blanco 1, Pleiades, and NGC 2516. We pick 21 clusters for testing: Blanco 1, Collinder 69, Huluwa 3, IC 4756, LP 2373 gp2, LP 2373 gp4, LP 2383, LP 2442, Mamajek 4, NGC 1980, NGC 2422, NGC 2451B, NGC 2516, NGC 3532, NGC 6405, NGC 6475, NGC 6633, Pleiades, Praesepe, Stephenson 1, and UBC 31. {We select these clusters from the \cite{pang65membership} catalog. These clusters are classified by five morphological categories: filamentary, fractal, halo, tidal tail and unspecified (i.e., without clear morphological features outside one tidal radius). The 21 cluster that we select contains clusters that belong to each of these categories (4 filamentary, 4 fractal, 3 halo, 5 tidal tail, and 5 unspecified clusters).} The data is retrieved from Astropy queries with full inclusion of the field stars. The membership labeling data is taken from \citet{pang65membership}. {Since the member stars from \citet{pang65membership} are incomplete for $m_G$ < $21$, we again exclude the dim stars with $m_G$ < $21$ in the {\it Gaia} DR3 data from our analysis. Note that this magnitude cut removes stars below 0.3 solar mass \citep{pang2024}, which does not affect our scientific motivation in this work. }

\subsection{Machine learning}

We test five ML algorithms: K-nearest neighbors \citep[KNN,][]{Cover1967NearestNP}, decision trees  \citep[DT,][]{BreiFrieStonOlsh84}, random forest  \citep[RF,][]{breiman2001random}, feed-forward neural network \citep[FFNN,][]{bebis1994feed}, and support vector machines  \citep[SVM,][]{cortes1995support}. 

KNN is a simple, effective, and widely used ML model. It compares the distances of the $K$ nearest neighbors to a given sample for making predictions \citep{peterson2009k}. The algorithm's performance depends on $K$. Previously, the algorithm was used on the {\it Gaia} data in combination with the GMMs for membership identification \citep{agarwal2021ml}. In our work, we use $K=5$. The distance between neighbors is measured in terms of Minkowski distance. The probability of a cluster membership is calculated based on the `distance' weight function. 

DT is a supervised ML algorithm that recursively splits the data based on features to create a tree-like structure \citep{destree}. We use a DT Classifier from Scikit-learn \citep{scikit-learn} with default parameters. To the best of our knowledge, this model has not yet been used in studies of OCs. 

RF is the ensemble learning ML algorithm that uses multiple decision tree predictors for classification and regression problems \citep{breiman2001random}. This method was used in several studies using the {\it Gaia} data \citep{gao2018machine, gao2018memberships, gao2019investigation, mahmudunnobe2021membership, das2023membership}. In our work, we configure the model with 100 trees, a `gini' tree split criterion, and without specification of depth of trees. We use an RF classifier from the Scikit-Learn library \citep{scikit-learn}.

The FFNN is the supervised ML model and is considered part of deep learning because the model may have depth depending on the number of layers \citep{bebis1994feed}. Selecting appropriate hyperparameters such as activation functions, optimization functions, learning rate, criterion, batch size, number of epochs, and number of layers is crucial for successful learning. In our case, we used 5-layer neural networks (NNs) with 400 batch size with activation functions such as LeakyReLU, Sigmoid, and GeLU and trained on 20 epochs with Adam optimizer and cross-entropy loss criterion. We use the PyTorch library \citep{paszke2017automatic} to implement FFNN. 

The SVM is a supervised machine learning algorithm that finds the optimal hyperplane that best separates different classes in the input feature space \citep{cortes1995support}. To our knowledge, it was not used in previous studies on OC membership. In this study, we configure SVM with default hyperparameters from Scikit-learn library \citep{scikit-learn}. 

For the evaluation criteria, we use the confusion matrix, which is a matrix that shows the number of true positives (TP), true negatives (TN), false positives (FP), and false negatives (FN) of binary classification, as shown in Table \ref{tab:matrix}. The diagonal part of the matrix is the correctly classified samples that are TN and TP \citep{kohl2012performance}. We measure the accuracy in terms of the $F_1$ score accuracy \citep{goutte2005probabilistic}:
\begin{equation}
F_1=\frac{2\times\mathrm{precision}\times\mathrm{recall}}{\mathrm{precision}+\mathrm{recall}},
\end{equation}
where
\begin{align}
    \mathrm{precision}=\frac{\mathrm{TP}}{\mathrm{TP}+\mathrm{FP}}, \ \ \  \  \mathrm{recall}=\frac{\mathrm{TP}}{\mathrm{TP}+\mathrm{FN}}.
\end{align}

\begin{table}
\begin{center}
\begin{tabular}{l|l|l|l|l}                                                                  
\multicolumn{2}{c}{}&\multicolumn{2}{c}{Prediction}&\\
\cline{3-4}
\multicolumn{2}{c|}{}&\multicolumn{1}{c|}{Non-Member}&\multicolumn{1}{c|}{Member}&\multicolumn{1}{c}{}\\
                                      
\hhline{~|---}
\multirow{2}{*}{Actual value}& Non-Member & TN & FP \\
                                      
\hhline{~|---}
& Member & FN & TP \\
                                      
\hhline{~|---}

\end{tabular}
\end{center}
\caption{Confusion matrix representing the performance of the prediction compared to the actual value.}
\label{tab:matrix}
\end{table}

We train our ML models on data from snapshots of $N$-body simulations and the {\it Gaia} DR3 data. Combining a few snapshots does not improve the performance. For the $N$-body simulation data, we consider snapshots with different SFEs at different time frames. For the default case, we use the snapshot at the end of violent relaxation at 20~Myr of the model cluster with SFE of $17\%$. We also consider snapshots at 100~Myr and {500~Myr, as discussed below.}

{For testing, we use snapshots from simulations with different SFEs. These simulations have randomization for positions and masses that are different from those of the training set.} When we add field stars, we use snapshots at 100 Myr and {1 Gyr.} An example testing set of model clusters with SFE of $17\%$ at 20~Myr snapshot combined with field stars can be seen in Fig.~\ref{fig:field_example}.

\section{Results}
\label{sec:results}

All the results shown below are obtained using the RF method. The dependence on ML methods is described in Appendix~\ref{ml_performance}. 

\subsection{Test on $N$-body simulation data}
\label{sec:nbody_result}

\begin{table}
    \centering
    \caption{Classification results of RF model when the Galactic field stars are included.}
    \begin{tabular}{|c|c|c|c|c|c|c|}
    \hline
    \textbf{SFE}&\textbf{Time} & \textbf{$F_1$} & \textbf{TP} & \textbf{TN} & \textbf{FP} & \textbf{FN}\\
    \textbf{[\%]} & [Myr] & [\%] &\# &\# &\# &\# \\
    \hline
    17&100 & 96.3& 2046&	11,114,147&	87& 74\\
    17&1000& 76.4& 99  &	11,419,272&	63&	1\\
    20&100 & 96.6& 3311&	11,024,986&	23&	214\\
    20&1000& 93.8& 650 &	10,976,115&	89&	0\\
    25&100 & 92.5& 5691&	10,660,405&	4&	920\\
    25&1000& 98.4& 3321&	10,822,245&	13&	98	\\
    \hline
    \end{tabular}
    \label{tab:field}
\end{table}

We first apply ML to $N$-body simulation data. For our first training, we select a snapshot at 20 Myr for a simulation with $17\%$ SFE. This snapshot corresponds to the system's phase of equilibrium after violent relaxation.

\begin{figure}
\centering
\includegraphics[width=0.8\columnwidth]{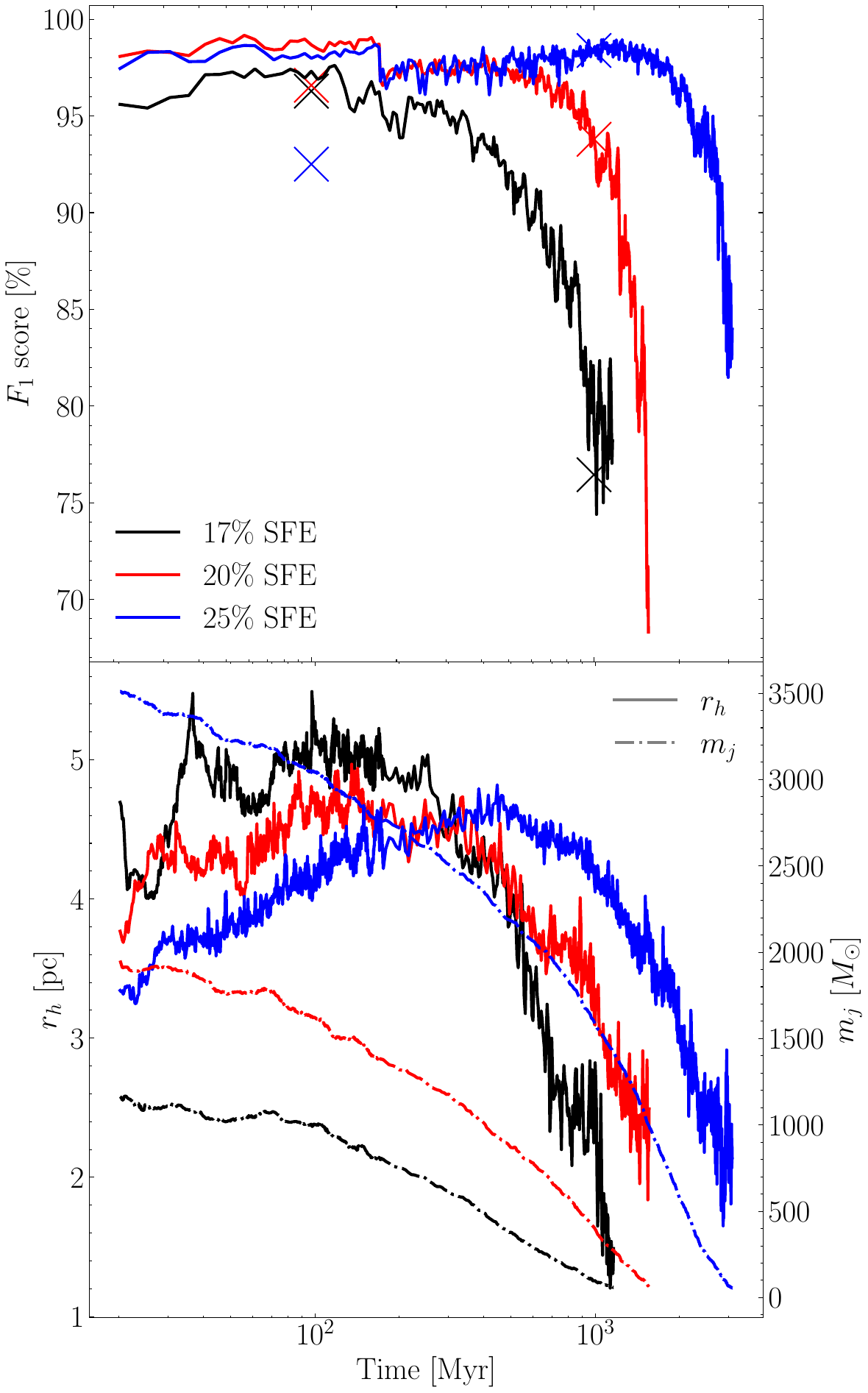}
\caption{$F_1$ score of ML models tested on $N$-body simulation data with different SFE as a function of time (top panel). The half-mass radius (solid lines) and Jacobi mass (dashed lines) are shown in the bottom panel.} 
\label{fig:basic}
\end{figure}

The upper panel of Fig. \ref{fig:basic} shows the $F_1$ score as a function of time for different testing sets. We first discuss the case with $17\%$ SFE, shown with black color. In the initial phase ($t\lesssim200 \, \mathrm{Myr}$), the accuracy is $\approx 95\%$. It starts to decrease with time at $t \sim 200 \, \mathrm{Myr}$, reaching $ \approx77.5\%$ at $t \sim 1 \, \mathrm{Gyr}$. The drop in accuracy is correlated with the size and mass of the cluster, which shrink over time, as seen in the bottom panel of Fig.~\ref{fig:basic}. The majority of the mistakes at later snapshots are due to an increase in FP classifications. 

Classification accuracy for simulations with higher SFE is qualitatively similar to $\mathrm{SFE}=17\%$ but with modest quantitative differences. The red and blue lines show $F_1$ score accuracy as a function of time for tests with $\mathrm{SFE}=20\%$ and $\mathrm{SFE}=25\%$. At $t\lesssim200 \, \mathrm{Myr}$, the accuracies are $\approx 98.0\%$ and $ \approx 97.5$ for $\mathrm{SFE}=20\%$ and $\mathrm{SFE}=25\%$, respectively. The $F_1$ scores start decreasing with time at $t \sim 1 \, \mathrm{Gyr}$ and  $t \sim 2 \, \mathrm{Gyr}$ for $\mathrm{SFE}=20\%$ and $\mathrm{SFE}=25\%$, respectively. The lowest value for $\mathrm{SFE}=20\%$ is $F_1\approx68\%$  at $t\approx1.557 \, \mathrm{Gyr}$ and for $\mathrm{SFE}=25\%$ it is $F_1\approx83\%$ at $t\approx3 \, \mathrm{Gyr}$.

\begin{figure}
\centering
\includegraphics[width=0.8\columnwidth]{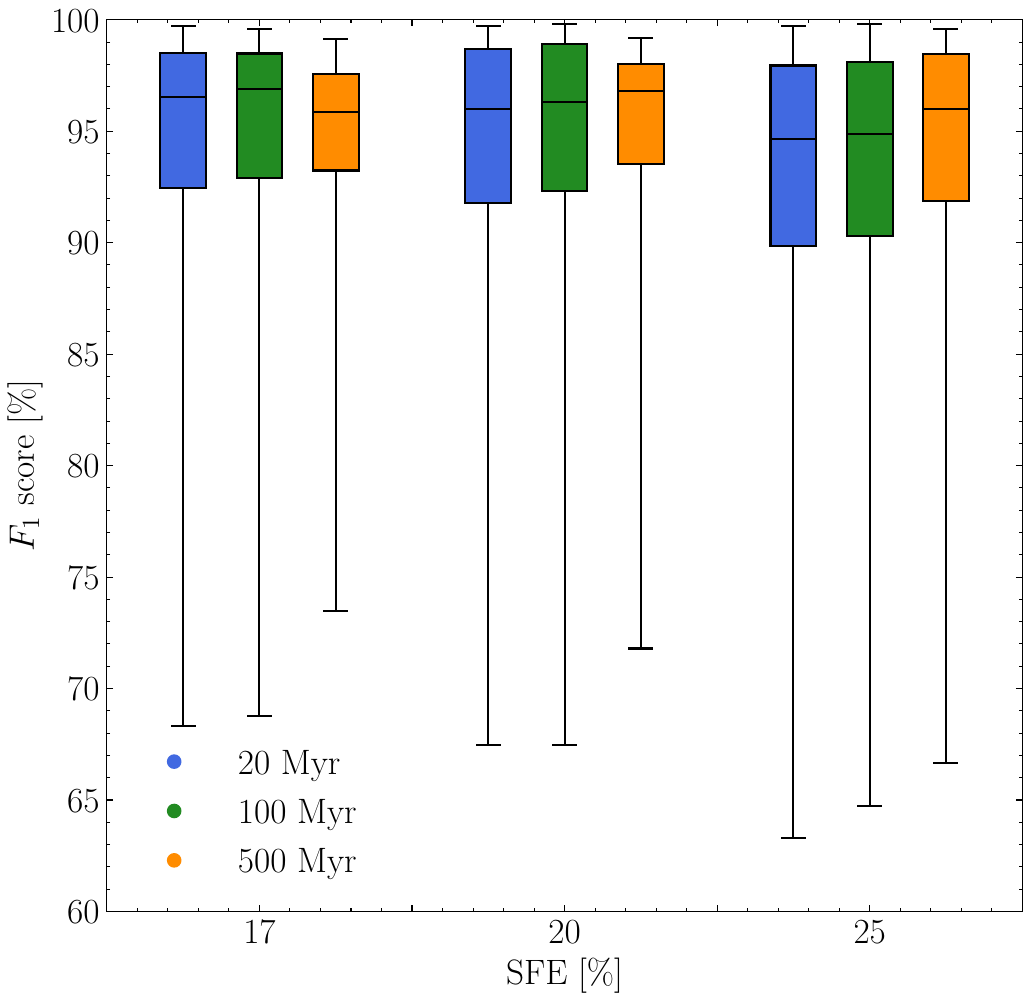}
\caption{Box plots of $F_1$ score as a function of SFE for different snapshots. The $x$-axis shows the SFE value of the training set. Color represents the snapshot time of training sets. Each box plot shows the median, quantiles, minimum, and maximum values of classification results on more than 1096 synthetic cluster datasets.}
\label{fig:trainingset}
\end{figure}

Next, we explore the dependence of training sets on the time and SFE of $N$-body simulation snapshots. We train nine models on snapshots of three model clusters with different SFEs (17, 20, and 25 \%) at three times (20, 100, and{ 500 Myr}). We test each model on 1096 snapshots with different times and SFEs representing the full snapshots of all $N$-body simulations used. Fig.~\ref{fig:trainingset} shows the $F_1$ score as box plots for different values of $t$ and SFE. The center of the box corresponds to the median $F_1$ score value, while the upper/lower boundaries correspond to $25\%$ quantiles. The error bars represent the max and min values. Overall, the differences in accuracies between different training sets are minor. The average accuracy varies between $F_1 \approx 95\%$ to $F_1 \approx 97\%$. The upper quantiles are between $ \approx 97\%$ and $ \approx 99\%$, and maximum values are $ \approx99\%$. The lower quantiles are between $89.0\%$ and $\approx 93.5\%$, and minimum values are between $63\%$ and $73\%$. We also train our models on a combination of up to $10$ snapshots, but we find that the resulting accuracy is not higher than that for models trained with only one snapshot. 

\begin{figure}
\centering
\includegraphics[width=0.8\columnwidth]{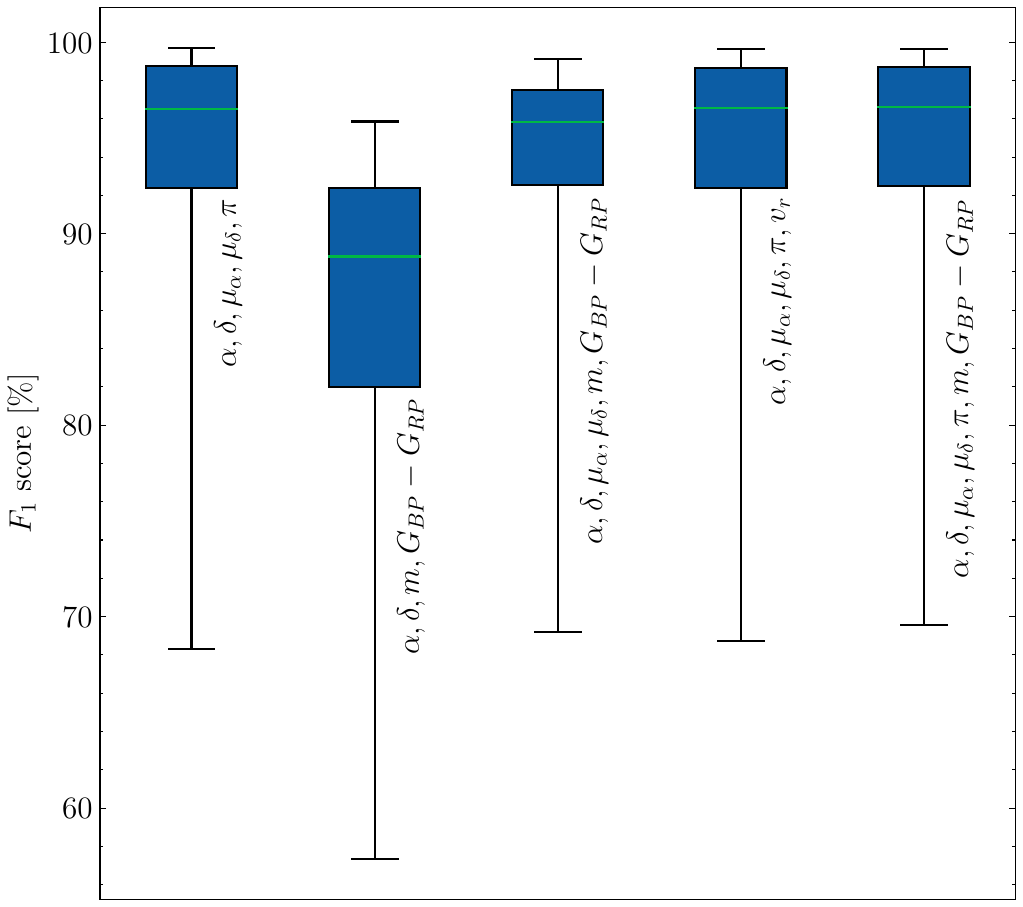}
\caption{Box plots of $F_1$ accuracy score for five different combinations of parameters used in the training set. Overall, the accuracy is not very sensitive to the combination of parameters used for training. The five-parameter combination ($\alpha$, $\delta$, $\mu_\alpha$, $\mu_\delta$, $\pi$) exhibits the highest accuracy of 96.5 \%. The four-parameter combination ($\alpha$, $\delta$, $m$, $G_{BP}-G_{RP}$) exhibits the lowest accuracy of 88.80\%.}
\label{fig:features_comparison}
\end{figure}

Next, we explore the impact of the stellar parameters on classification accuracy. As described above, we use five different combinations of parameters (see Table \ref{tab:feature}). Fig.~\ref{fig:features_comparison} shows the $F_1$ score for these five combinations. The five-parameter combination ($\alpha$, $\delta$, $\mu_\alpha$, $\mu_\delta$, $\pi$) exhibits the highest median accuracy of $\approx 96.5\%$. The upper and lower quantiles are $ \approx 92.4\%$ and $\approx 98.7\%$. The four-parameter combination ($\alpha$, $\delta$, $m$, $G_{BP}-G_{RP}$) exhibits the lowest median accuracy of $\approx88.8\%$. The upper and lower quantiles are $\approx 82.0\%$ and $\approx 92.4\%$. The remaining combinations have results similar to that of the five-parameter combination ($\alpha$, $\delta$, $\mu_\alpha$, $\mu_\delta$, $\pi$) in medians, quantiles and min-max values.    

We now explore the impact of field stars. We add field stars as non-members to the training set at 20 Myr. We then test the model on 100 Myr and 1 Gyr snapshots. The corresponding $F_1$ scores are shown as crosses in the top panel of Fig.~\ref{fig:basic}. The inclusion of the field does not affect the $F_1$ score significantly. The accuracy of the test at $t=1\, \mathrm{Gyr}$ matches that of without field stars. For the tests at 100 Myr, all the accuracies are slightly lower compared to results without field stars. For $\mathrm{SFE}=17\%$ and $\mathrm{SFE}=20\%$, the accuracies are 1.1\% and 2.0\% lower. For $\mathrm{SFE}=25\%$, the drop in accuracy is 5.5\%. The values are provided in Table \ref{tab:field}.

\subsection{Tests on observational data}
\label{sec:obs_results}

\begin{figure} 
\begin{center}
\centering
\includegraphics[width=0.9\columnwidth]{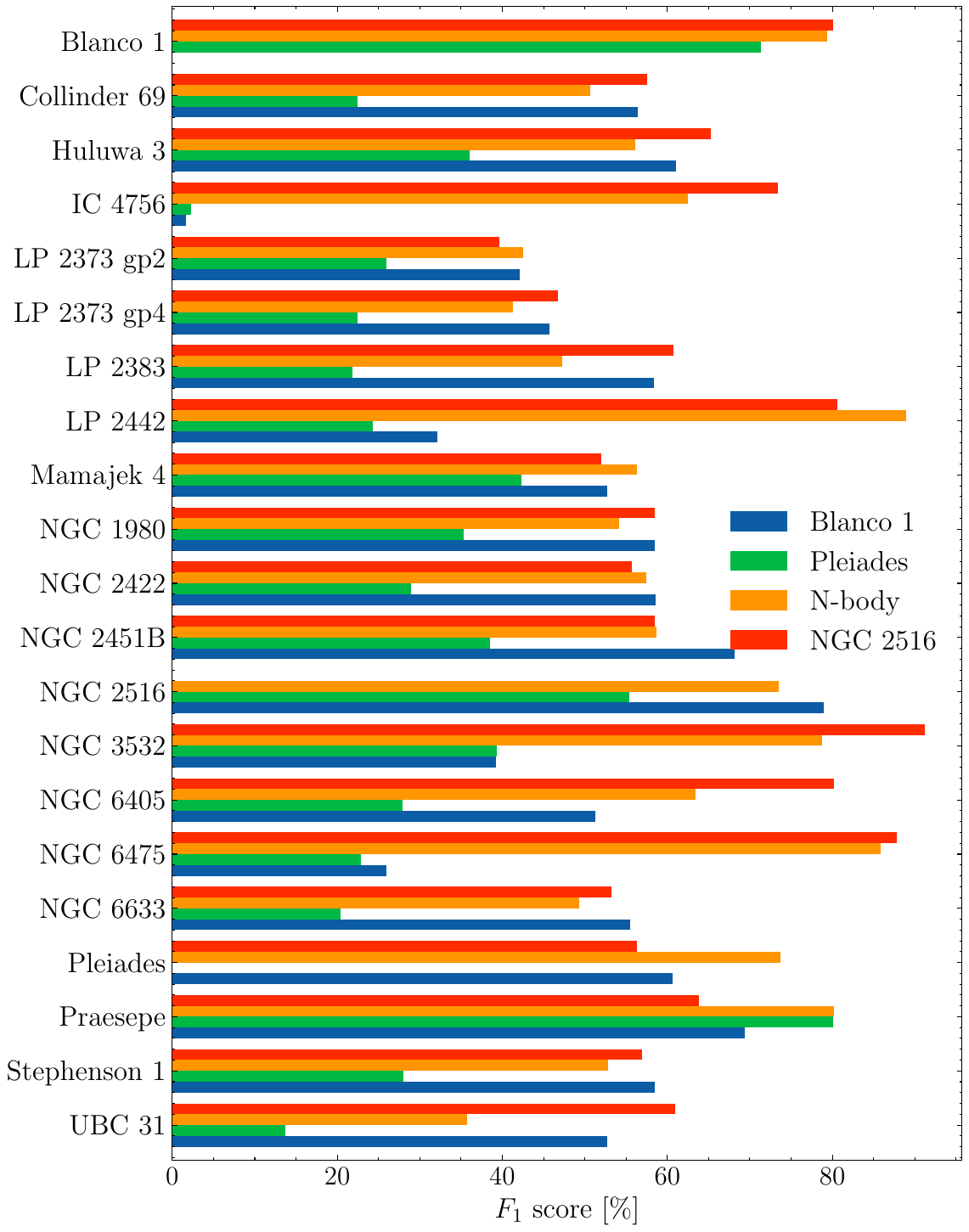}
\caption{$F_1$ score accuracies for 21 different clusters. Instances with the same training and testing set are dropped.} 
\label{fig:obs_overall}
\end{center}
\end{figure}

We now apply our ML approach to the clusters from the {\it Gaia} data. As mentioned, we train on four different models (default set of $N$-body simulation, Blanco 1, Pleiades, and NGC 2516 clusters) and then test on 21 {\it Gaia} clusters. Fig.~\ref{fig:obs_overall} shows the $F_1$ score for all training and testing data combinations. In the following, we first discuss the impact of training data and then the impact of the testing data. 

The models trained on $N$-body simulation, Blanco 1, Pleiades, and NGC 2516 clusters yield average $F_1$ scores\footnote{Here, we average the $F_1$ scores of all tested {\it Gaia} clusters for each model. When we average, we exclude the instances where the training set is the same as the testing data.} of 61.4, 51.4, 33.0, and 60.8\%, as summarised in Table~\ref{tab:train_char}. The highest accuracy of $\approx 91.1\%$ is seen when the model trained on NGC 2516 is tested on NGC 3532. The lowest accuracy of $F_1\leqslant1\%$ is seen for models trained on Blanco 1, and Pleiades are tested on IC 4756. Field stars are misclassified as 22296 and 11408 FPs for these two models.

We now analyze the impact of the quantitative properties of the training sets on classification accuracy. Table \ref{tab:train_char} provides the mean $F_1$ scores averaged over {\it Gaia} data for models trained on $N$-body simulation, Blanco 1, Pleiades, and NGC 2516 clusters. We do not see a clear dependence between accuracy and the training sets' mass, age, and $r_h$. However, the mean $F_1$ score tends to be higher in the models trained on clusters with larger member stars, but the dependence is not monotonic.

\begin{table}[]
    \centering
    \begin{tabular}{|c|c|c|c|c|c|}
    \hline
    \textbf{Cluster} & \textbf{Members} & \textbf{Mass} & \textbf{Age} &\textbf{$r_h$} &\textbf{Mean $F_1$}\\
    & \#& [$M_\odot$] & [Myr] & [pc] & [\%]  \\
    \hline
    Blanco 1 & 703 & 338.6 & 100 & 6.7 & 51.4\\
    \hline
    Pleiades & 1407 & 740.6 & 125 & 4.7 & 33.0\\
    \hline
    $N$-body & 2139 & 1164.2 & 25&4.7 & 61.4\\
    \hline
    NGC 2516 & 2690 & 1984.8 & 123& 7.9 & 60.8\\
    \hline
    \end{tabular}
    \caption{Characteristics of training sets and the values of $F_1$ scores. Columns from 1 to 5 show the cluster name, number of member stars, mass, age, and half-mass radius $r_h$. The mean $F_1$ score (last column) is the average value of $F_1$ scores obtained by testing on the {\it Gaia} data.} 
    \label{tab:train_char}
\end{table}

\begin{figure*} 
\begin{center}
\centering
\includegraphics[width=1.8\columnwidth]{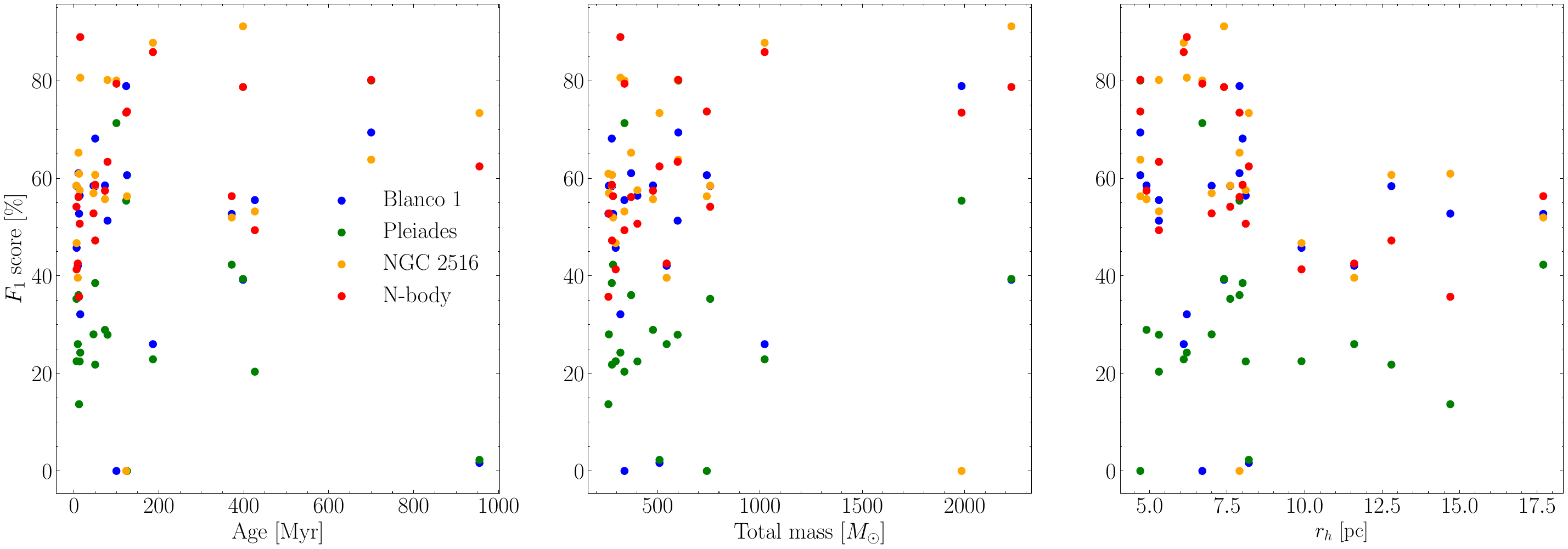}
\caption{$F_1$ score as a function of age, mass, half mass radius, and density. Different colors correspond to models trained with Blanco 1, Pleiades, NGC 2516, and $N$-body simulations. Individual dots correspond to the result of testing on clusters from the {\it Gaia} data.} 
\label{fig:mass_f1}
\end{center}
\end{figure*}

Figure \ref{fig:mass_f1} shows the scatter plots of $F_1$ score for different clusters as a function of the age (left panel), total mass (middle panel), and half-mass radius (right panel). We do not observe any clear correlation between the $F_1$ score and these quantities. The model trained on the Pleiades cluster exhibits overall lower accuracy than other models. This is consistent with the lowest average $F_1$ score of 31.4\% mentioned above.  

\begin{figure} 
\begin{center}
\centering
\includegraphics[width=\columnwidth]{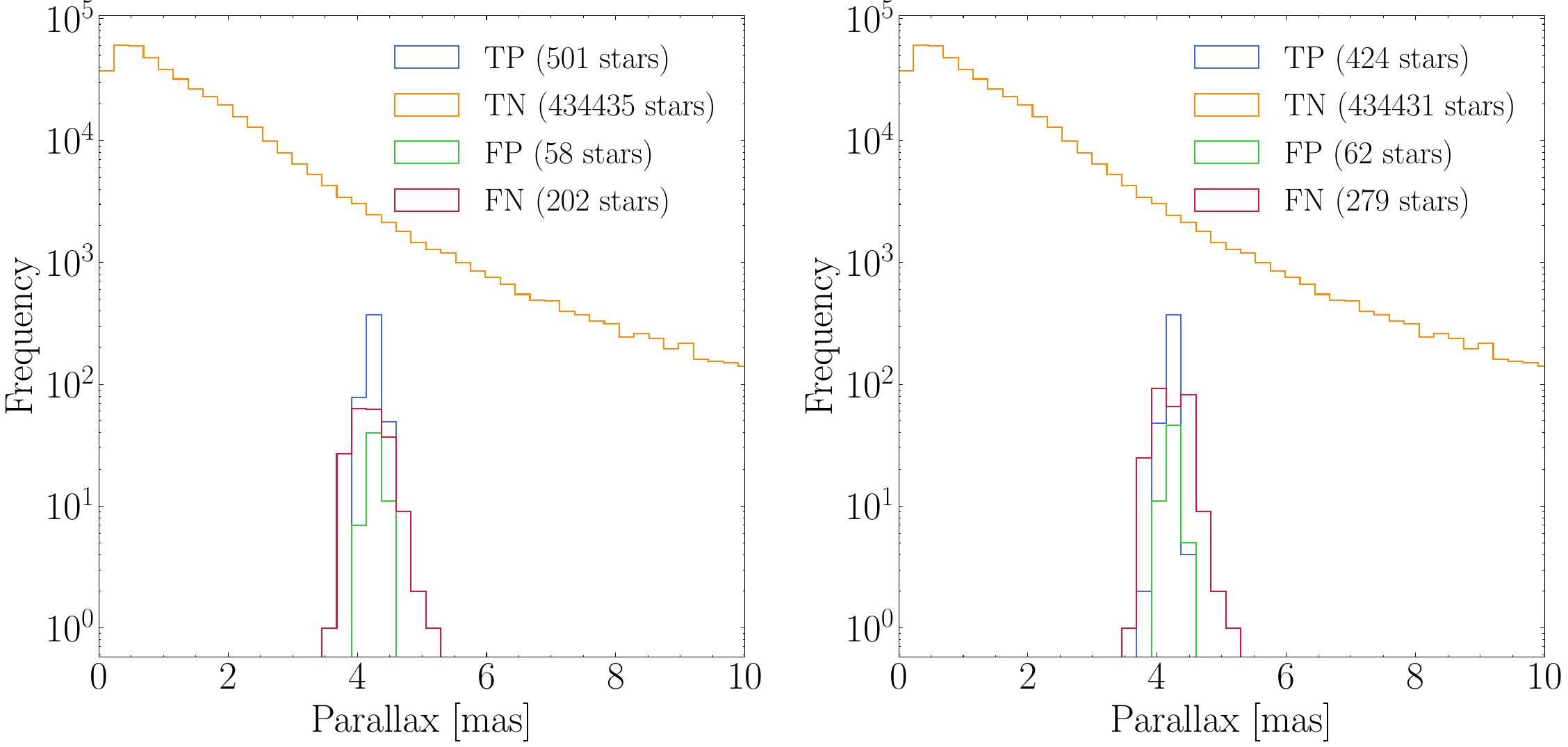}
\caption{Classification results in terms of parallax for the models trained on $N$-body simulation (left panels) and Pleiades (right panel) and tested on the Praesepe cluster.}
\label{fig:nbody_vs_pleiades_parallax}
\end{center}
\end{figure}

As an example of classification, Fig.~\ref{fig:nbody_vs_pleiades_parallax} shows the histograms of parallaxes for the Blanco 1 cluster. We apply the model trained on $N$-body simulation (left panel) and Pleiades (right panel). The blue, orange, green, and red colors represent the histograms of TPs, TNs, FPs, and FNs. Overall, the two methodologies yield similar results. The accuracy is $\approx 79.0\%$ and $\approx 71.0\%$ for models trained on $N$-body simulations and Pleiades. For the former model, the number of TPs is 501, while that of FNs is 202. For the latter model, we have 424 TPs and 279 FNs.

\begin{figure} 
\begin{center}
\centering
\includegraphics[width=0.95\columnwidth]{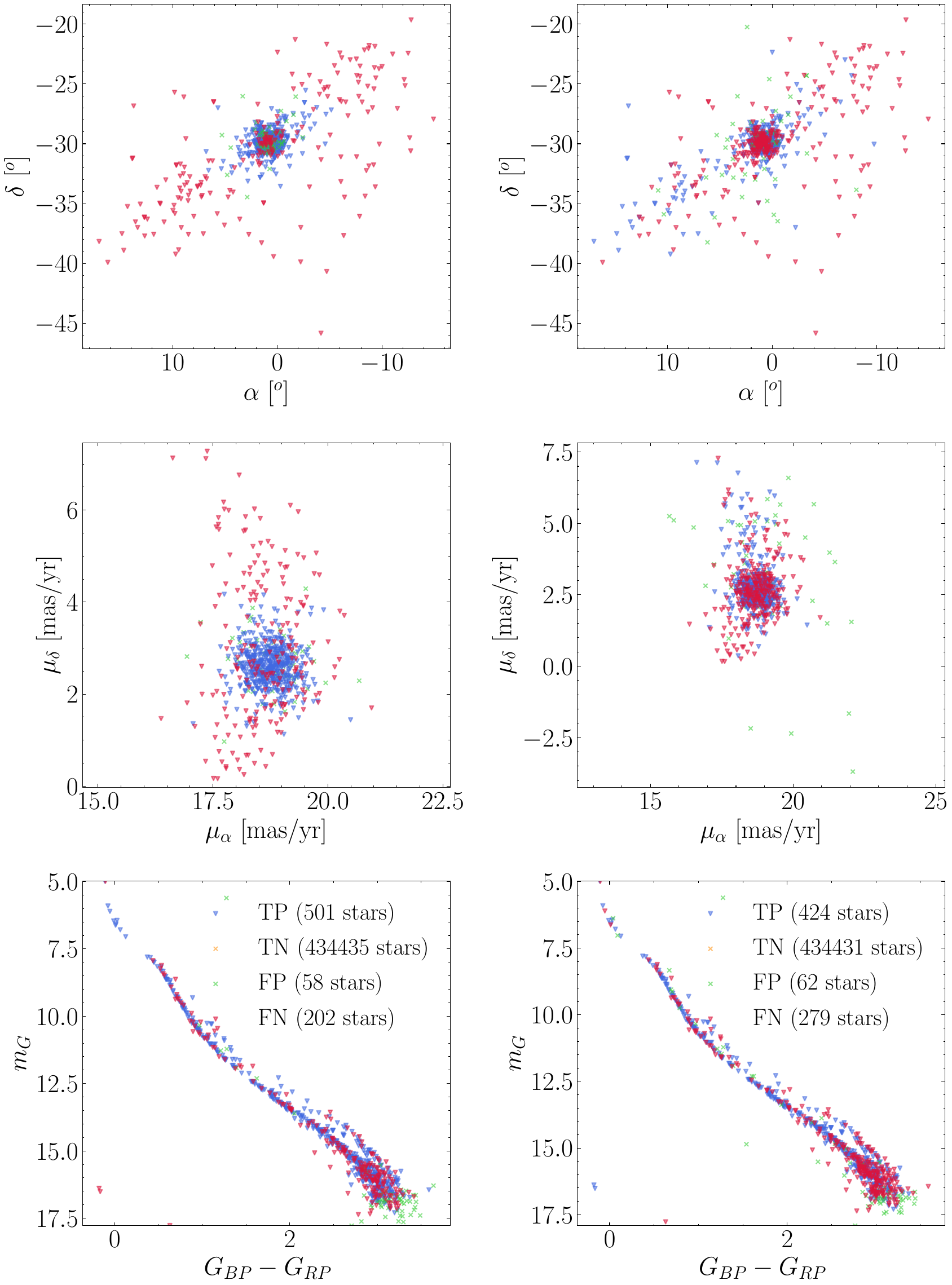}
\caption{Classification results for individual stars for the models trained on $N$-body simulation (left panels) and Pleiades (right panel) and tested on the Praesepe cluster. The top panel shows coordinates $\alpha$ and $\delta$. Middle panels show proper motions in $\alpha$ and $\delta$. The bottom panels show apparent magnitude and color. Due to their large number, the true negatives are not visualized in these plots.}
\label{fig:nbody_vs_pleiades}
\end{center}
\end{figure}

The reason behind this behavior can be identified by looking at individual stars. Figure~\ref{fig:nbody_vs_pleiades} shows the FPs, FNs, and TPs with green, red, and blue dots for individual stars as a function of coordinates $\alpha$ and $\delta$ (top panels), proper motions in $\alpha$ and $\delta$ (middle panels), and magnitudes (bottom panels). The left panel is for a model trained on $N$-body simulation, and the right panel is for a model trained on the Pleiades cluster. The TP stars of the model trained on the $N$-body appear closer to the center than the TPs of the model trained on the Pleiades. This is visible both in the top and middle panels. The locations of FP stars are different in the top and middle panels. FPs are mainly located near the cluster center for the model trained on $N$-body simulations. However, the FPs are primarily located outside the cluster for the model trained on the Pleiades. FNs appear in similar locations, but there are more FNs for models trained on the Pleiades inside the cluster in the top and middle panels. As we can see in the bottom panels, most FPs and FNs of both models are low-mass (dim and cold) stars in the plot's bottom right part. 

\subsection{Comparison with other methods}
\label{sec: trad_compare}

We compare { the membership derived from} our approach to { that from} non-machine learning method of \cite{rosen_praesepe}, which is based on a modified convergent-point method. { At the same time, we also make a comparison to membership from \cite{pang65membership} that used the unsupervised machine learning method \texttt{StarGo}}. { we choose} the Praesepe cluster {as an example to carry out the comparison}. We use stars within the same field of view as in the previous sections (cf. Section~\ref{sec:method}). Among 1393 { member} stars from \cite{rosen_praesepe}, 1079 are in this { field of view}. The {stars outside the selected field of view are considered as the} tidal tail of Praesepe. 

\begin{figure}
    \centering
    \includegraphics[width=\columnwidth]{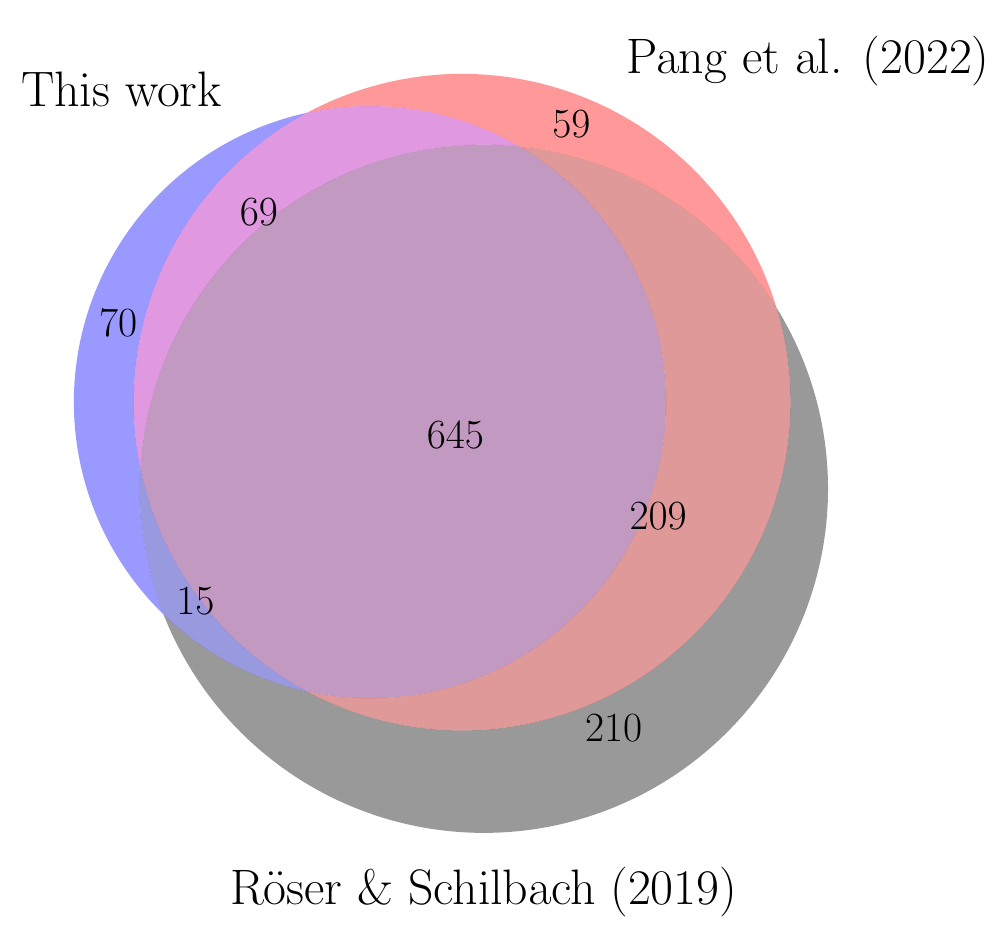}
    \caption{Venn diagram of comparison of three membership identification methods. Blue corresponds to our method, while gray and red correspond to \cite{rosen_praesepe} and \cite{pang65membership}, respectively.}
    \label{fig:venn}
\end{figure}

 We display the Ven diagram of the comparison of the three methods in Fig.~\ref{fig:venn}. The blue circle corresponds to our method, while the red and gray circles represent the members of \cite{pang65membership} and \cite{rosen_praesepe}, respectively. Our method identifies 799 stars as members of Praesepe, while \cite{pang65membership} and \cite{rosen_praesepe} identified 982 and 1079 member stars, respectively. Among all stars, 645 are cross-matched in all three memberships. Seventy stars are identified as members that are not recognized by the other two methods.
 Two hundred ten members identified in \cite{rosen_praesepe} are not recovered in the other two methods, most of which are located in the extended tidal tails. 

\begin{figure*}
    \centering
    \includegraphics[width=1.8\columnwidth]{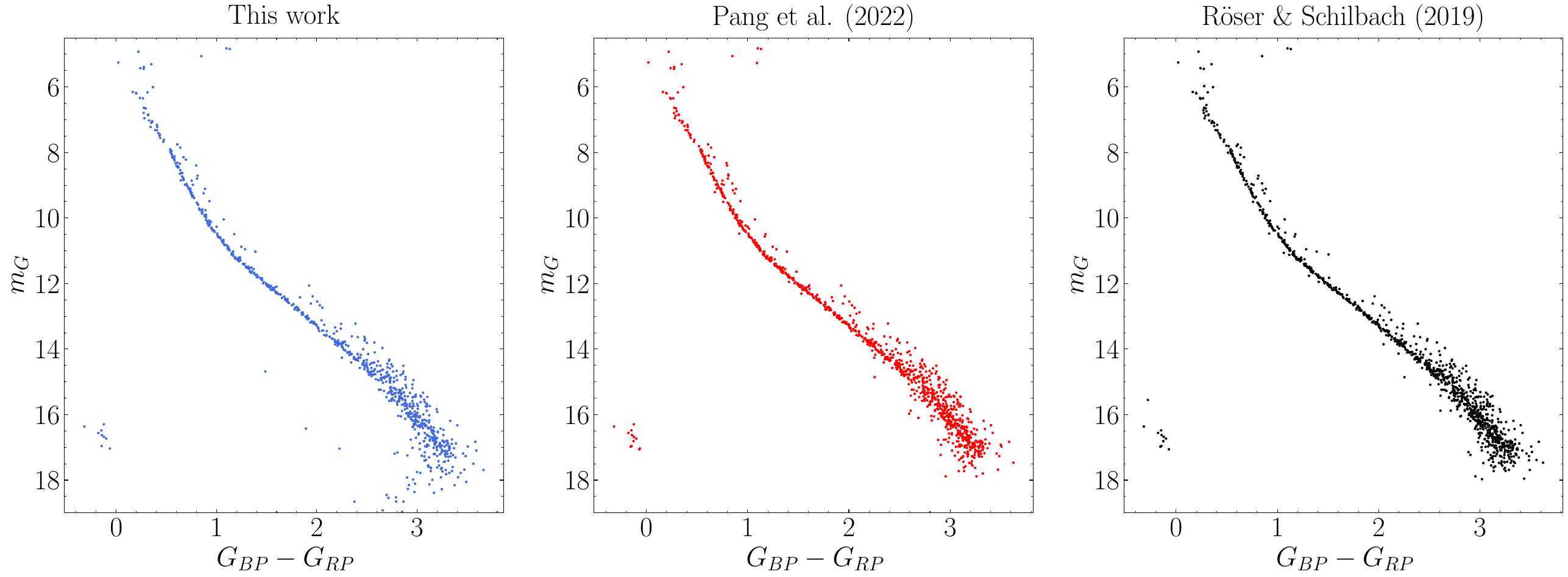}
    \caption{Comparison of three membership identifications by the color-magnitude diagrams.}
    \label{fig:trad_compare}
\end{figure*}

Fig \ref{fig:trad_compare} shows color-magnitude diagrams (CMDs) for the member stars identified by these three methods. All three show similar { patterns}. { Our method yields a larger number of faint stars with low effective temperatures. Because we do not apply quality cut to the stars before membership identification. These faint stars have larger observational uncertainties than the majority of stars.}

\begin{figure*}
    \centering
    \includegraphics[width=1.8\columnwidth]{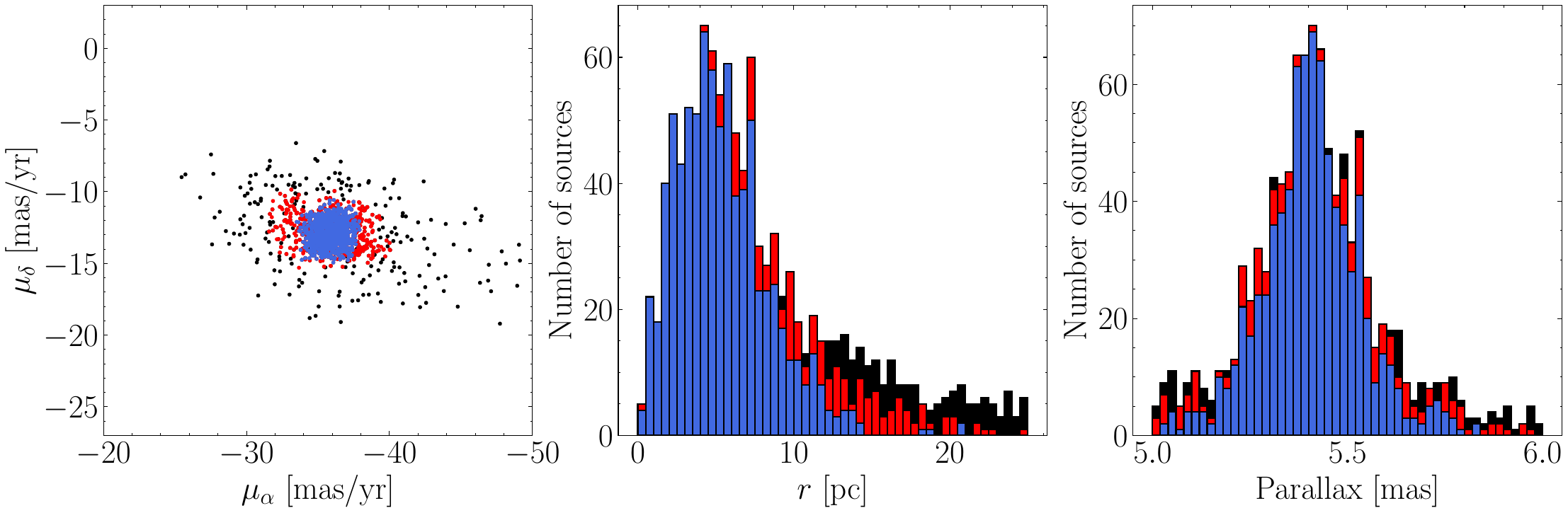}
    \caption{Comparison of 3 membership labels in terms of proper motions in $\alpha$ and $\delta$ (left panel), radial distribution (middle panel), parallaxes (right panel). Blue, red, and black represent member stars identified by our method, \cite{pang65membership} and \cite{rosen_praesepe}, respectively.}
    \label{fig:trad_compare_other}
\end{figure*}

The left panel of Fig. \ref{fig:trad_compare_other} { displays} the distribution of stars by the proper motions in $\alpha$ and $\delta$. { Members from our method are very concentrated} around $\mu_\delta \sim -12 $ and $\mu_\alpha \sim 35$ $\mathrm{mass}/\mathrm{yr}$ within a radius of $\sim 4 \, \mathrm{mass}/\mathrm{yr}$.{ Members of} \cite{pang65membership} { are more extended than} this region { roughly} within a radius of $\sim 5 \, \mathrm{mass}/\mathrm{yr}$. \cite{rosen_praesepe}'s { members are distributed most extended}, well outside ($\gtrsim 10 \, \mathrm{mass}/\mathrm{yr}$) the proper motion center. This is consistent with the radial distribution shown in the middle panel of Fig.~\ref{fig:trad_compare_other}: 4 member stars identified by our method are located further than 15 pc from the cluster center, while \cite{pang65membership} and \cite{rosen_praesepe} memberships have 73 and 194 stars more than 15 pc away from the cluster center. Similarly, {the parallax distribution (right panel of Fig.~\ref{fig:trad_compare_other}} reveals that the member star identified by our method is confined to a central concentrated region. The main reason for this difference stems from the fact that our method uses proper motion and parallax for training.

\section{Conclusion}
\label{sec:conclusion}

In this work, we perform membership analysis of stellar clusters using supervised machine learning { algorithm}. We train and test our models on snapshots { data} from $N$-body simulations of stellar clusters and observed clusters from the {\it Gaia} DR3 data. Our findings can be summarised as follows. 

We { study} five supervised ML algorithms on $N$-body simulation data. These models are Random Forest (RF), Decision Trees (DT), Support Vector Machines (SVM), Feed-Forward Neural Networks (FFNN), and K-Nearest Neighbors (KNN). All models produce comparable accuracies within $\sim 1\%$ (see Section~\ref{sec:a1}). Following this result, we use the RF method for the rest of the paper. 

We then { explore} the impact of eight different observational parameters on the { accuracy of membership identification}. Among these eight, five are astrometric parameters: right ascension ($\alpha$), declination ($\delta$), proper motions in $\alpha$ ($\mu_\alpha$) and $\delta$ ($\mu_\delta$), and parallax ($\pi$); two are photometric parameters: apparent magnitude $m$ and color parameter $G_{BP}-G_{RP}$; and the radial velocity, which is measured with spectroscopy. We tested five different combinations of these eight parameters. The highest accuracy was observed in models trained on purely astrometric parameters $\alpha, \delta, \mu_\alpha, \mu_\delta, \pi$. { The parallax is the most critical parameter in member classification.} The inclusion of $r_v$ and photometric features { seem not} improve accuracy. { However, this might be biased by the current observational uncertainty of $r_v$.}

We { study} the impact of star formation efficiency and the cluster's age on the classification accuracy using $N$-body simulation data. The $F_1$ score accuracy, { which defines the reliability of our ML method}, is $\gtrsim 90\%$ in all snapshots before cluster dissolution. When the cluster is mainly dissolved, $F_1$ score accuracy drops to $\sim70\%$. The errors in classification during the dissolution are primarily due to false negatives. Statistical results on three snapshots of different SFEs at different times show that the accuracy is similar regardless of the time and SFE of the snapshot used for training. 

Additionally, we explore the impact of the number of member and non-member stars in datasets. Generally, for the successful performance of the supervised ML models, it is desirable to have a similar amount of samples within all the classes. Datasets of open clusters typically contain a large number of field stars, which are not members. We found that the number of field stars within the training dataset does not affect the classification accuracy. 

We { apply} our model to 21 clusters from the {\it Gaia} DR3 data. These clusters are Blanco 1, Collinder 69, Huluwa 3, IC 4756, LP 2373 gp2, LP 2373 gp4, LP 2383, LP 2442, Mamajek 4, NGC 1980, NGC 2422, NGC 2451B, NGC 2516, NGC 3532, NGC 6405, NGC 6475, NGC 6633, Pleiades, Praesepe, Stephenson 1, and UBC 31. In addition to $N$-body simulation data, we train our model on Blanco 1, Pleiades, and NGC 2516 clusters. The models trained on $N$-body simulation and NGC 2516 yield an average $F_1$ of $\approx60\%$. The models trained on Blanco 1 and Pleiades show a lower average $F_1$ score of $\approx51\%$ and $\approx32\%$. The models trained on clusters (trained on NGC 2516 and N-body simulation) with a larger number of member stars tend to yield higher classification accuracy, but the dependence is not monotonic. {The two models with the highest accuracy both have over 2000 member stars in the training set. In contrast, the two models with the lowest accuracy (trained on Blanco 1 and Pleiades) have fewer than 2000 members.} Among these clusters, we find no noticeable correlation between the classification accuracy and the mass, age, and half-mass radius ($r_h$) of clusters.

We compare our membership determination results with the membership of \cite{pang65membership} and \cite{rosen_praesepe} for the Praesepe cluster. In total, 645 member stars are cross-matched in all three methods. Our model retrieves 124 and 280 fewer stars than \cite{pang2020different} and \cite{rosen_praesepe}, respectively. The members identified by our method are more concentrated in both space distribution and proper motion distribution compared to the other two methods. Our models are trained on five-parameter space: right accession, declination, parallax, and proper motions. Therefore, parallax and proper motions play roles that place a larger weight on member identification.

Our work suggests that machine learning approaches are promising in membership analysis, regardless of several limitations. In our work, we used a limited number of 21 {\it Gaia} clusters. The inclusion of more clusters should improve the accuracy. It is also worthwhile comparing the unsupervised learning approach with other machine learning methods (e.g., StarGO, DBSCAN, HDBSCAN, Gaussian Mixture Models). Also, cross-matching the results with other studies on membership analysis can further improve results. These limitations will be addressed in future studies. 

%%%%%%%%%%%%%%%%%%%%%%%%%%%%%%%%%%%%%%%%%%%%%%%%%%
\section{Data Availability}

The data used in this study can be obtained from the authors upon request.

%%%%%%%%%%%%%%%%%%%%%%%%%%%%%%%%%%%%%%%%%%%%%%%%%%%%%%%%%%%%%%%%%%%%%
\begin{acknowledgements}
%%%%%%%%%%%%%%%%%%%%%%%%%%%%%%%%%%%%%%%%%%%%%%%%%%%%%%%%%%%%%%%%%%%%%
{The authors would like to thank the anonymous referee for his/her valuable comments and suggestions, which have helped to improve the quality of this manuscript.}
This research has been funded by the Science Committee of the Ministry of Education and Science, Republic of Kazakhstan (Grant No. AP13067834, AP19677351, and Program No. BR20280974). Additionally, funding is provided through the Nazarbayev University Faculty Development Competitive Research Grant Program, with Grant No. 11022021FD2912. Xiaoying Pang acknowledges the financial support of the National Natural Science Foundation of China through grants 12173029 and 12233013. 
Peter Berczik thanks the support from the special program of the Polish Academy of Sciences and the US National Academy of Sciences under the Long-term program to support the Ukrainian research teams grant No.~PAN.BFB.S.BWZ.329.022.2023.
\end{acknowledgements}

\bibliographystyle{aa.bst}
\bibliography{new} 

\begin{appendix} 

\section{Dependence on machine learning model}
\label{sec:a1}

In this section, we check how our results depend on the ML model. We apply Random Forest (RF), Decision Trees (DT), Support Vector Machines (SVM), Feed-Forward Neural Networks (FFNN), and K-Nearest Neighbors (KNN) models.

\begin{figure}
\centering
\includegraphics[width=0.7\linewidth]{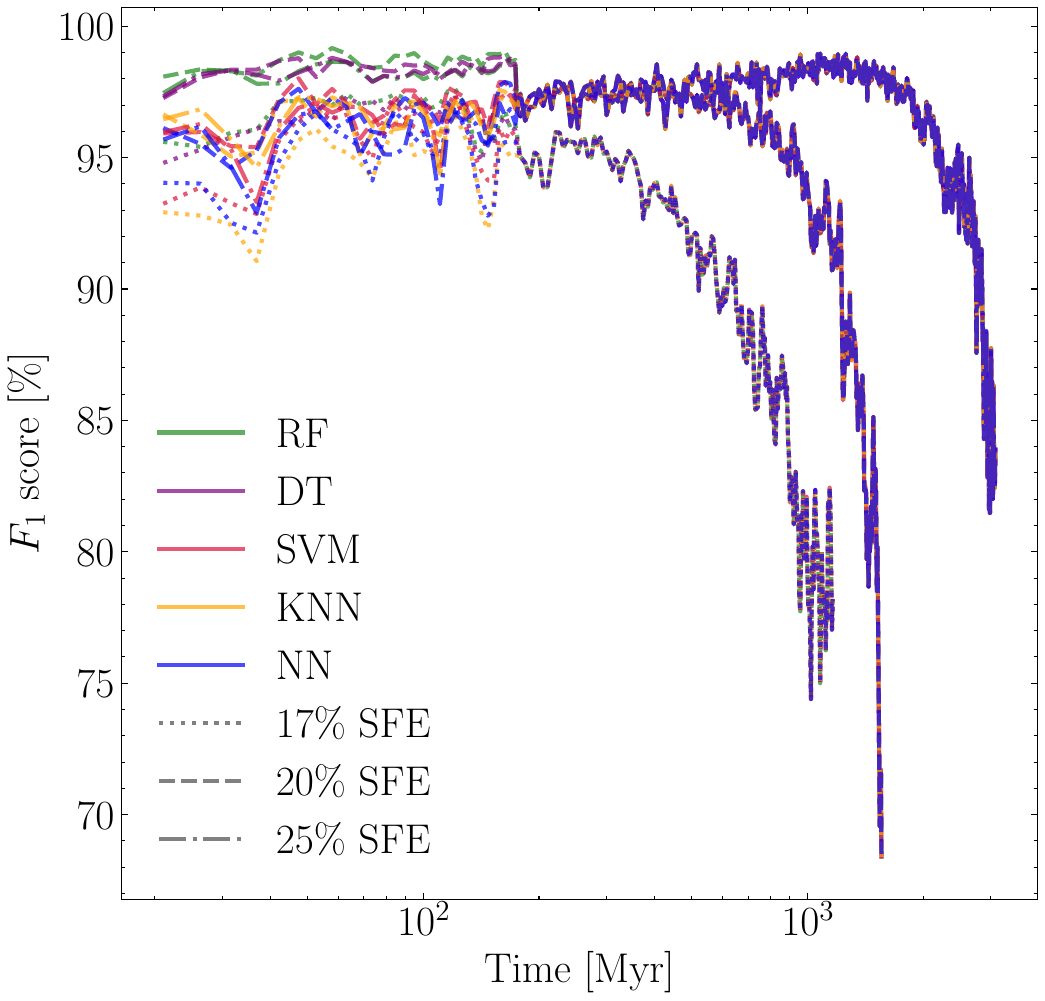}
\caption{Average $F_1$ score as a function of time for different ML models. The training is performed on $N$-body simulation snapshots at 20 Myr.}
\label{fig:ML_comparison}
\end{figure}

Figure~\ref{fig:ML_comparison} shows the $F_1$ score as a function of time for different values of SFE and different ML models. The green, purple, red, orange, and blue lines represent $F_1$ score accuracy as a function of time for RF, DT, SVM, FFNN, and KNN models. All ML models exhibit similar values of the $F_1$ score. The difference between different models does not exceed $\sim 1\%$. The RF model performs slightly better within this difference, especially for $t \lesssim 200$ Myr. We thus adopt the RF model in the main part of our work. 

\label{ml_performance}

\end{appendix}

\end{document}